\documentclass[reprint,amsmath,amssymb,pra,longbibliography]{revtex4-2}

\usepackage{graphicx}% Include figure files
\usepackage{color}
\usepackage[caption=false]{subfig}

\usepackage{bm}% bold math

\newcommand{\vare}{\varepsilon}
\newcommand{\RE}{\mathrm{Re}}
\newcommand{\IM}{\mathrm{Im}}

\begin{document}

% \preprint{APS/123-QED}

\title{Parametric dependence of bound states in the
	continuum: a general theory}

\author{Amgad Abdrabou}
\email{abdrabou@zju.edu.cn}
\affiliation{School of Mathematical Sciences, Zhejiang University, Hangzhou 310027,
	China}

\author{Lijun Yuan}
\affiliation{School of Mathematics and Statistics, Chongqing Technology and Business University, Chongqing 400067, China}

 \author{Wangtao Lu}
 \affiliation{School of Mathematical Sciences, Zhejiang University, Hangzhou 310027,
 	China }
 
\author{Ya Yan Lu}
\email{Corresponding author: mayylu@cityu.edu.hk}
\affiliation{Department of Mathematics, City University of Hong Kong,
	Kowloon, Hong Kong, China}

\date{\today} 
\begin{abstract}
  Photonic structures with high-$Q$ resonances are essential for many
  practical applications,  and they can be relatively easily realized by
  modifying ideal structures with bound states in the continuum
  (BICs). When an ideal photonic structure with a BIC is perturbed,
  the BIC may be destroyed (becomes a resonant state) or may continue to exist with 
  a slightly different  frequency and a slightly different wavevector (if
  appropriate). Some BICs are robust against certain structural
  perturbations, but most BICs are nonrobust. Recent studies suggest
  that a nonnegative integer $n$ can be defined for any generic
  nondegenerate BIC with respect to a properly defined set of structural
  perturbations. The integer $n$ is the minimum number of tunable
  parameters needed to preserve the BIC for perturbations arbitrarily 
  chosen from the set. Robust and nonrobust BICs have $n=0$ and $n\ge
  1$, respectively. A larger $n$ implies that the BIC is more
  difficult to find. If a structure is 
  given by $m$ real parameters, the integer $n$ is the 
  codimension of a geometric object formed by the parameter
  values at which the BIC exists in the $m$-dimensional parameter
  space. In this paper, we 
  suggest  a formula for $n$, give some justification for the general
  case,  calculate $n$ for different types of BICs in two-dimensional 
  structures with a single periodic direction, and illustrate the
  results by numerical examples. Our study improves the theoretical
  understanding on BICs and provides useful guidance to their
  practical applications. 
\end{abstract}
\maketitle

\section{Introduction}
\label{S1}

In an open wave system, a bound state in the continuum (BIC) is an
eigenmode with a localized wave field (localized in the open spatial directions) and a
frequency inside the radiation continuum~\cite{hsu16,sad21}. An
ideal BIC can be regarded as a resonant state with an infinite quality
factor ($Q$-factor)~\cite{hsu13}, and it gives rise to high-$Q$ resonances when the
structure and/or solution parameters (such as the wavevector) are
perturbed~\cite{bistab17,hu18,perturb18,kosh18,perturb20}. In recent years, many
applications of BICs have been realized in
photonics~\cite{kosh19,azzam}. 
Most of these applications rely on field
enhancement~\cite{yoon15,moca15,hu20_1,lfe22} or sharp 
features in scattering spectra~\cite{shipman05,gipp05,bykov15,blan16}
caused by the BIC-induced   
high-$Q$ resonances. From a theoretical point of view, it is important
to understand how BICs are formed and what properties they
possess. Existing studies have identified a number of physical
mechanisms under which BICs can be found~\cite{hsu16,sad21}. Rigorous proofs for the existence of BICs
are available for symmetry-protected BICs~\cite{evans94,bonnet94,shipman12}. In
structures that are extended to infinity in one or two spatial
directions, a BIC is characterized by its frequency and wavevector. It
is known that some special BICs are surrounded by resonances whose
$Q$-factors diverge with unusually high rates~\cite{perturb18,perturb20,jin19}. 
Moreover, BICs in a periodic structure are polarization
singularities in momentum space and they can be 
characterized by a topological charge defined using the polarization
vector of the surrounding resonant states~\cite{zhen14,bulg17pra,fudan,notomi,amgad21}. 

Another important theoretical question is concerned with the effect of
structural perturbations on BICs. For a symmetry-protected BIC, it is
clear that the BIC should still exist if the perturbation preserves the
symmetry. This means that the BIC is robust with respect to
symmetry-preserving perturbations. On the other hand, for a typical
symmetry-breaking perturbation of magnitude $\delta$, the BIC
is destroyed and becomes a resonant state with an $Q$-factor proportional
to $1/\delta^2$~\cite{kosh18}. However, for special symmetry-breaking perturbations,
the BIC may become a resonant state  with a larger
$Q$-factor (proportional to $1/\delta^4$, or $1/\delta^6$, ...)~\cite{perturb20}, and
it can even remain as a BIC. Some BICs
unprotected by symmetry are 
also robust with respect to certain structural perturbations.
As examples, we mention propagating BICs in periodic structures 
with both the up-down mirror symmetry and the in-plane inversion
symmetry~\cite{yuan17_4,robust21}, and BICs in 
optical waveguides with lateral leakage channels~\cite{robustoe}. Even though these
BICs are unprotected by symmetry, their robust existence still
depends crucially on symmetry.
Similar to the case of symmetry-protected BICs, if the structural
perturbation breaks the required symmetry, the BIC is usually but not
always destroyed. 

In recent works~\cite{para20,para21}, a new approach was developed to
characterize nonrobust BICs. The idea is to determine
the minimum number of tuning parameters needed to continuously follow
the BIC. More precisely, 
assuming a BIC exists in a structure with dielectric function
$\varepsilon_*({\bm r})$ where ${\bm r}=(x,y,z)$ is the position vector, we
try to find the smallest integer $n$, so that the BIC persists in 
the perturbed structure with dielectric function
\begin{equation}
  \label{perturbed}
  \varepsilon({\bm r})
  = \varepsilon_*({\bm r}) + \delta F({\bm r})+ \gamma_1 G_1({\bm r}) + ... +
  \gamma_n G_n({\bm r}), 
\end{equation}
where $F$, $G_1$, ..., $G_n$ are arbitrary perturbation
profiles, $\delta$ is an arbitrary small real number, $\gamma_1$, ...,
$\gamma_n$ are tuning parameters determined together with
the BIC. If $n=0$, i.e., no tuning parameters are necessary, then the BIC is
robust. If $n \ge 1$, the BIC is nonrobust. A larger $n$ implies that the BIC is more
difficult to find.
The number $n$ also describes how the BIC depends on
generic structural parameters~\cite{para21}. If the structure depends on $m$
parameters, we can consider the geometric object  $P_{\rm BIC}$ formed by the parameter
values at which the BIC exists in the $m$-dimensional parameter
space, then $n = m - \dim (P_{\rm BIC})$ is the codimension of $P_{\rm
  BIC}$. For example,  if $n=1$, the BIC forms a curve in the plane
of two parameters and a surface in the 3D space of
three parameters.

However, the existing studies have covered only the cases $n=1$
and $n=2$ for some BICs in 2D structures with a single periodic
direction~\cite{para20,para21}. In this paper, we present a
general theory for generic
nondegenerate BICs.
The rest of this paper is organized as follows. In Sec.~\ref{S2}, we
present a general theory including a formula for $n$ and a brief
justification. In Sec.~\ref{S3},  we apply the general theory to BICs
in 2D structures that are translationally invariant in one spatial
direction and periodic in another direction. Some details are given
for cases that have not been analyzed in  previous
works~\cite{para20,para21}. Numerical examples for the new cases
of Sec.~\ref{S3} are presented in Sec.~\ref{S4}. The paper is concluded with a brief 
discussion in Sec.~\ref{S5}. 

\section{General theory}
\label{S2}

To present the theory in a more concrete setting, we assume an
electromagnetic BIC of frequency $\omega_*$ exists in a lossless
non-magnetic structure given 
by scalar dielectric function $\vare_*({\bm r})$.
The structure may be translationally
invariant  and/or periodic in one or two directions, thus the BIC may possess a wavevector ${\bm \alpha}_*$ which contains at most two nonzero
components. The electric field of the BIC can be written as
\begin{equation}
  \label{gBloch}
  {\bm E}_* ({\bm r}) = {\bm \Phi}_*({\bm r}) e^{ i  {\bm \alpha}_*
    \cdot {\bm  r}} 
  \end{equation}
  where ${\bm \Phi}_*$ has the same invariant and/or periodic
  directions as $\vare_*$. For example, if the structure is
  a photonic crystal  (PhC) slab with 2D periodicity in the  $xy$-plane, then
  ${\bm \alpha}_* = (\alpha_*, \beta_*, 0)$, where $\alpha_*$ and
  $\beta_*$ are the Bloch wavenumbers.

  The definition of BIC requires the existence of at least one
  radiation channel. This implies that the structure must have at
  least one open direction where outgoing and incoming waves with the
  same frequency $\omega_*$ and wavevectors compatible with ${\bm
    \alpha}_*$ can propagate to and from infinity, respectively. For a PhC
  slab surrounded by air, the $z$ variable (perpendicular to the slab)
  provides the open directions (as $z \to \pm \infty$), and the
  propagating diffraction orders (compatible with ${\bm \alpha}_*$) serve as the radiation channels. For each radiation
  channel, we have at least one and at most two linearly independent 
  scattering solutions. The case of two solutions in one radiation
  channel arises when the 
  incident waves can have different polarizations. Let us denote the
  electric fields of the independent scattering solutions by ${\bm
    E}_k^{(s)} ({\bm r}) = {\bm \Psi}_k({\bm r}) e^{ i {\bm \alpha}_* \cdot {\bm
      r}}$ for $k=1$, 2, ...,  $N_{\rm ss}$, where $N_{\rm ss}$ is
  the total number of such solutions.

 In order to discuss the robustness of a BIC and determine the
 codimension $n$ of nonrobust BICs, we must specify the conditions
 (most importantly, the symmetries) satisfied by the perturbation
 profiles $F({\bm r})$ and $G_j({\bm   r})$, $1\le j \le n$. The original
 structure, given by the dielectric function $\vare_*({\bm r})$, may have more symmetries than the perturbation
profiles. For example, a PhC slab may have $C_4$ or $C_6$ symmetry,
but robustness can be discussed for perturbations with $C_2$ symmetry
only, where $C_p$ is the rotation by $2\pi/p$ about the $z$-axis. The
symmetries in the original structure and the perturbations give rise
to related symmetries in the BIC and the scattering
solutions. Typically, this requires a proper scaling. 
For example, in a  PhC slab with $C_2$ symmetry [same as
the in-plane inversion symmetry, namely $\vare({\bm r}) =
\vare(-x,-y,z)$], the BIC and the  
scattering solutions can be scaled such that
\begin{equation}
  \label{gPT}
  \begin{bmatrix}
  \overline{E}_x({\bm r})  \cr
    \overline{E}_y({\bm r}) \cr
    \overline{E}_z({\bm r})  
  \end{bmatrix}
  =
  \begin{bmatrix}
    E_x (-x, -y, z) \cr
    E_y (-x, -y, z) \cr
    -E_z (-x, -y, z)
  \end{bmatrix},
\end{equation}
where $E_x$ is the $x$-component of ${\bm E}_*$ or ${\bm E}^{(s)}_k$,
$\overline{E}_x$ is its complex conjugate, etc~\cite{robust21}. Notice that ${\bm
  \Phi}_*$ and ${\bm \Psi}_k$ also satisfy Eq.~(\ref{gPT}). 

For each scattering solution, we define an integer index. 
First, we consider the linear operator ${\cal L}$, such that 
the BIC and the scattering solutions (with $e^{i {\bm \alpha}_* \cdot
  {\bm   r}}$ removed) satisfy 
\begin{equation}
  \label{gopL}
  {\cal L} {\bm \Phi}_* = {\bm 0}, \quad {\cal L}{\bm \Psi}_k = {\bm     0}, 
\end{equation}
for $k=1$,  ..., $N_{\rm ss}$. The inhomogeneous equation
\begin{equation}
  \label{ginho}
  {\cal L} {\bm u} = {\bm f}
\end{equation}
usually has outgoing solutions, since ${\bm f}$ in the right hand side
serves as a source. If we insist that Eq.~(\ref{ginho}) has a solution
that decays to zero rapidly in the open 
direction(s), then 
\begin{equation}
  \label{gScond}
  \int_\Omega \overline{ \bm \Psi}_k \cdot {\bm f} \, d{\bm r}
  =   \int_\Omega \overline{ \bm \Psi}_k \cdot {\cal L}  {\bm u} \, d{\bm
    r}
  =
  \int_\Omega \overline{  {\cal L} \bm \Psi_k } \cdot  {\bm u} \, d{\bm
    r}   = 0, 
\end{equation}
where $\Omega$ covers one period in the periodic
direction and a unit length in the invariant direction. The index $I_k$
is defined as the number of real constraints on ${\bm f}$ for the left
hand side above being zero. Since ${\bm \Psi}_k$ is complex, we
usually have $I_k=2$. However, if ${\bm f}$ has some useful symmetry
related to the symmetry of the BIC and the perturbation profiles,
$I_k$ may be reduced to $1$ or $0$.
% Thus,  $I_j \in \{0, 1, 2 \}$ can only be
% determined by considering the specific symmetry of the BIC and the
% perturbations.
Moreover, if there are two scattering solutions  in the same radiation
channel, they should be chosen so that the constraints are independent.

If a BIC exists in the perturbed structure with the dielectric function given in
Eq.~(\ref{perturbed}),  it should have a  frequency near
$\omega_*$ and a wavevector near ${\bm  \alpha}_*$. The degrees of
freedom $N_{\rm wv}$ of the BIC wavevector, is the number of
components in ${\bm \alpha}_*$ that will vary independently. Usually
$N_{\rm wv}=2$, but if the BIC is generic and 
${\bm \alpha}_* = {\bm 0}$, then, as we will show in Sec.~\ref{S3}, the perturbed BIC also has a zero 
wavevector, thus $N_{\rm wv}=0$. If ${\bm \alpha}_*$ has just one nonzero
component, the wavevector  of the perturbed BIC 
may have one or two nonzero components depending on the symmetry, thus
$N_{\rm wv}$ can be 1 or 2, respectively.

Based on the above definitions, we believe that the integer $n$ of a 
generic nondegenerate BIC is
\begin{equation}
  \label{formula}
  n = \sum_{k=1}^{N_{\rm ss}} I_k - N_{\rm wv}.
\end{equation}
To find $n$, we actually need to show that a BIC exists in the
perturbed structure with the minimum of $n$ tunable parameters. To do
so, we expand the desired BIC [with frequency 
$\omega$, wavevector ${\bm \alpha}$, electric field
${\bm E}({\bm r})= {\bm \Phi}({\bm r}) e^{ i {\bm \alpha} \cdot {\bm r}}$] and the
tunable parameters in power series of $\delta$:
\begin{eqnarray}
  \label{gExpand}
  && {\bm \Phi} = {\bm \Phi}_* + \delta {\bm \Phi}_1 + \delta^2 {\bm 
     \Phi}_2 + \cdots, \\
  &&  \omega = \omega_* + \delta \omega_1 + \delta^2 \omega_2 + \cdots, \\  
  && {\bm \alpha} = {\bm \alpha}_* + \delta {\bm \alpha}_1 + \delta^2
     {\bm \alpha}_2 + \cdots, \\  
  \label{gExpgama}
  &&   \gamma_j  = \delta \gamma_{j,1}  + \delta^2 \gamma_{j,2} +
     \cdots, \quad j=1, ..., n.
\end{eqnarray}
The leading order gives ${\cal L}{\bm \Phi}_* = {\bm 0}$, i.e., the
original equation satisfied by the BIC. For $l \ge 1$, the ${\cal
  O}(\delta^l)$ equation can be written as 
\begin{equation}
  \label{gIter}
  {\cal L} {\bm \Phi}_l = {\bm  f}_l  
\end{equation}
where ${\bm f}_l$ is related to $\omega_l$, ${\bm \alpha}_l$,
$\gamma_{j,l}$, ${\bm \Phi}_*$, ..., ${\bm \Phi}_{l-1}$, $\vare_*$ and the
perturbation profiles. 

If Eq.~(\ref{gIter}) has a solution
that decays rapidly in the open direction(s) and has the same
symmetry as the original BIC, we obtain
$\sum_{k=1}^{N_{\rm ss}}  I_k$ constraints on ${\bm f}_l$ from the $N_{\rm ss}$
scattering solutions. In addition, similar to Eq.~(\ref{gScond}), we have
\begin{equation}
  \label{biccond}
  \int_\Omega \overline{\bm \Phi}_* \cdot {\bm f}_l \, d{\bm r}
  = 0.  
\end{equation}
The above always gives one real constraint. Therefore, we have the
total of $1 + \sum_{j=1}^{N_{\rm ss}} I_j$ real equations for solving
$\omega_l$, ${\bm \alpha}_l$ and $\gamma_{j,l}$ for $1\le j \le n$. The
total number of real unknowns is $1 + N_{\rm wv} + n$. We choose  $n$
to satisfy Eq.~(\ref{formula}), so that the total number of
equations is exactly the total number of unknowns. Since the unknowns
appear in ${\bm f}_l$ linearly with coefficients only related to
$\vare_*$, ${\bm \Phi}_*$, $F$ and $G_j$, we have a real square matrix
$A$ and a real vector ${\bm b}_l$, such that 
\begin{equation}
  \label{gMatrix}
  A
  \begin{bmatrix}
    \omega_l \cr
    {\bm \alpha}_l \cr 
    \gamma_{1,l} \cr
    \vdots \cr
    \gamma_{n,l}
  \end{bmatrix}
  = {\bm b}_l, 
\end{equation}
where ${\bm  \alpha}_l$ includes only $N_{\rm wv}$ components. The matrix $A$
depends on the BIC, the scattering solutions and the perturbation
profiles, but it does not depend on the previous iterations ${\bm
  \Phi}_s$ for $1 \le s < l$. Importantly, Eq.~(\ref{gScond}), with
${\bm f}$ replaced by ${\bm f}_l$, and Eq.~(\ref{biccond})  are sufficient conditions for
Eq.~(\ref{gExpand}) to 
have a solution that decays in the open direction(s) and preserves the
symmetry of the original BIC~\cite{para21,robust21}. Therefore, if $A$ is invertible, we can 
solve $\omega_l$, ${\bm \alpha}_l$ and $\gamma_{j,l}$ from 
Eq.~(\ref{gMatrix}), then solve ${\bm \Phi}_l$ from 
Eq.~(\ref{gIter}). This implies that we can iteratively find the BIC
and the tuning parameters through the power series. 

The theory is applicable to generic BICs defined as those for which the matrix $A$ is
invertible. The perturbation profiles should be arbitrary except for
some specified conditions (such as the symmetry). Even for a generic BIC, if the
perturbations are improperly chosen, the matrix $A$ can be
non-invertible, then a BIC may not exist in the perturbed
structure. On the other hand, there are also non-generic BICs for
which the matrix $A$ is always non-invertible for any choice of the
perturbation profiles.

\section{BICs in 2D periodic structures}
\label{S3}

In this section, we consider  BICs in 2D lossless dielectric structures that are invariant
in $x$, periodic in $y$ with period $L$, bounded in $z$ by $|z| <
d$,  and surrounded by air. The dielectric function $\vare(y,z)$ of such
a structure is real and satisfies 
\begin{eqnarray}
  \label{periodic} 
  && \vare(y+L,z) = \vare(y,z),   \quad \forall \ (y,z) \in
     \mathbb{R}^2, \\
  \label{air}
  && \vare(y,z) = 1, \quad |z| > d.
\end{eqnarray}
BICs in 2D structure with 1D periodicity have been
investigated by many authors~\cite{shipman03,port05,mari08,bulg14,hu15,yuan17,hu20pra}. Very often, one assumes that the structure
has the following additional symmetry:
\begin{eqnarray}
\label{C2} 
  && \vare(y,z) = \vare(-y,z), \quad \forall (y,z) \in\mathbb{R}^2, \\
\label{Sz} 
  && \vare(y,z) = \vare(y,-z), \quad \forall (y,z) \in\mathbb{R}^2. 
\end{eqnarray}
In recent works~\cite{para20,para21}, the codimension $n$ for 
some BICs in 2D periodic structures with the up-down mirror symmetry,
i.e., Eq.~(\ref{Sz}), has been determined. In the following, we apply the general
theory of Sec.~\ref{S2} to all cases with or without the symmetry
conditions (\ref{C2}) and (\ref{Sz}).

Let $\vare_*(y,z)$ be the dielectric function of a specific periodic
structure [satisfying Eqs.~(\ref{periodic}) and (\ref{air})], in which
there is a nondegenerate BIC with frequency $\omega_*$ and
wavevector ${\bm \alpha}_* = (\alpha_*, \beta_*, 0)$, where $\alpha_*$
and $\beta_*$ are real wavenumbers in the $x$ and $y$
directions, respectively. Due to the periodicity in $y$, we can assume 
$\beta_* \in (-\pi/L, \pi/L]$. For simplicity, we consider BICs with
$\omega_*$ satisfying 
\begin{equation}
  \label{1rad}
  \sqrt{ \alpha_*^2 + \beta_*^2} < \frac{\omega_*}{c}
  < \sqrt{ \alpha_*^2 + \left( \frac{2\pi}{L} - |\beta_*| \right)^2}, 
\end{equation}
where $c$ is the speed of light in vacuum. The above condition implies that only the zeroth
diffraction order is propagating and all other diffraction orders
(corresponding to the $y$-wavenumber $\beta_* + 2 m \pi/L$ for $m 
 \ne 0$) are evanescent. 

To apply the theory of Sec.~\ref{S2}, we first consider the
degrees of freedom $N_{\rm wv}$ for the BIC wavevector. 
If both $\alpha_*$ and
$\beta_*$ are nonzero, a BIC in the perturbed structure should have
wavenumbers $\alpha$ near $\alpha_*$ and $\beta$ near $\beta_*$, and
thus $N_{\rm wv}=2$. If $\alpha_* \ne 0$ and $\beta_* = 0$, we claim
that the perturbed BIC must have $\beta=0$ and thus $N_{\rm wv} = 1$. 
This is true only when the BIC in the unperturbed structure is
generic, so that there is only one BIC (near the original one) can be
found in the perturbed structure. Recall that a BIC is a special point in 
a band of  resonant states with a dispersion relation $\omega =
 \omega_\dagger(\alpha, \beta)$, where $\omega_\dagger$ is a
 complex-valued function of $\alpha$ and $\beta$. Due to 
 reciprocity and the reflection symmetry in $x$, $\omega_\dagger$
 satisfies 
 \begin{eqnarray}
   \label{recip}
   &&  \omega_\dagger(\alpha, \beta) = \omega_\dagger(-\alpha,
      -\beta),  \\
   \label{refle}
   &&  \omega_\dagger(\alpha, \beta) = \omega_\dagger(-\alpha, \beta).
 \end{eqnarray}
 The above two equations lead to $\omega_\dagger(\alpha, \beta) =\omega_\dagger(\alpha, -\beta)$. Therefore, if there is a BIC in the perturbed structure with a small nonzero
$\beta$, there must be another one with wavenumber $-\beta$.
  Since the original BIC is generic, this is not possible. Similarly, if the
original BIC has $\alpha_* = 0$ and $\beta_* \ne 0$,
then the perturbed BIC must also have $\alpha=0$, and thus $N_{\rm
  wv}=1$. Finally, if the unperturbed structure has a generic BIC with
$\alpha_*=\beta_*=0$, then the perturbed BIC also have
$\alpha=\beta=0$, and $N_{\rm wv}=0$.

In Table~\ref{tab1},
\begin{table}[h]
  \caption{Codimension of BICs in 2D periodic structures with different symmetry.}
  \centering
  \begin{tabular}{|c||c|c|c|c|} \hline
    $(\alpha_*, \beta_*)$ & $(\square,\square)$ &  $(\square,0)$ & $(0,\square)$ & $(0,0)$\\ \hline \hline
   No symmetry & 6 & 3 & 3 & 2 \\ \hline
   $y\leftrightarrow -y$ & 2 & 1 & 1 & 0 \\ \hline
   $z\leftrightarrow -z$ & 2 & 1 & 1 & 1 \\ \hline    
   $y\leftrightarrow -y$, $z\leftrightarrow -z$ & 0 & 0 & 0 & 0 \\ \hline    
  \end{tabular}
  \label{tab1}
\end{table}
we list the codimension $n$ for different kinds of BICs according to the symmetry of the structure and the zero pattern of the wavevector.  
In the first column of Table~\ref{tab1}, the reflection symmetries in $y$ and $z$ are denoted
as $y \leftrightarrow -y$ and $z \leftrightarrow -z$, respectively.
The first row shows the zero pattern of the wavevector $(\alpha_*, \beta_*)$, where 
$\square$ denotes a nonzero entry.
The last two rows summarize the results already obtained in previous
works~\cite{para20,para21}. 
In the following, we give some justification for the new results
listed in the table.

First, we consider the case of no symmetry.  The perturbation
profiles $F$ and $G_j$ for $1\le j\le n$, do not need to satisfy
Eq.~(\ref{C2}) or (\ref{Sz}), but they must be real functions of $y$
and $z$, periodic in $y$ with period $L$, and vanish for $|z| > d$.
Since the BIC in the unperturbed structure satisfies condition
(\ref{1rad}), there are two radiation channels (below and above the
periodic layer, respectively) corresponding to the zeroth diffraction
order. 

For a generic BIC with $\alpha_* \ne 0$ and $\beta_* \ne 0$, we need
to consider both polarizations in each radiation channel. Thus, the
total number of independent scattering solutions is $N_{\rm ss} =
4$. The constraint for each scattering solution is a complex
condition, thus the index for each scattering solution is $I_k =
2$. Therefore, the codimension of the BIC is
$n = 4 \times 2 - 2 = 6$.

If $\alpha_* \ne 0$ and $\beta_* = 0$, the BIC propagates along the
$x$ axis, has a vectorial field ${\bm E}_* = {\bm \Phi}_*  e^{  i
  \alpha_* x}$ with ${\bm \Phi}_*$ depending on $y$ and
$z$ only. Moreover, the BIC can be scaled such that the $x$ component of ${\bm
  \Phi}_*$ is pure  imaginary and the $y$ and $z$ components of
${\bm \Phi}_*$ are real~\cite{para21}. Similarly, we can scale the scattering
solutions ${\bm E}_k^{(s)} = {\bm \Psi}_k e^{ i \alpha_* x}$, so that 
the $y$ and $z$ components of ${\bm \Psi}_k$ are real and the 
$x$ component of ${\bm \Psi}_k$ is pure imaginary. As a result, the integral
condition $\int_\Omega {\bm \Psi}_k \cdot {\bm f}_l \, d{\bm r} = 0$
gives one real equation, and $I_k = 1$. Since $\Omega$ is a 3D
domain with a unit length in $x$, the above integral on
$\Omega$ is identical to the integral on the 2D cross section of
$\Omega$ given by $0 <
y < L$ and $-\infty < z < \infty$. Since $N_{\rm ss}=4$ and $N_{\rm wv}=1$, we
have $n = 4\times 1 - 1 = 3$. 

The BIC with $\alpha_*=0$ and $\beta_* \ne 0$ is a scalar mode in
 the $E$ or $H$ polarization. The electric or magnetic field has only
 one nonzero component (the $x$ component). Since the fields for
 opposite polarizations are 
orthogonal, we can consider only the scattering solutions in the
same polarization as the BIC. Therefore, $N_{\rm ss}=2$. Since $I_k = 2$
and $N_{\rm wv}=1$, we have $n=2\times 2 -1 = 3$.

If the BIC is a standing wave with $\alpha_* = \beta_* = 0$, we have
$I_k = 1$ and $N_{\rm ss}=2$ for the same reasons given above. Therefore,
$n = 2\times 1 - 0 =2$.

Next, we consider the case with reflection symmetry in $y$.
Notice that we assume $\vare_*$, $F$ and $G_j$ all satisfy symmetry
condition (\ref{C2}). In addition, $F$ and $G_j$ must be real and
periodic as before.  Since the structure is
invariant in $x$, Eq.~(\ref{C2})  is identical to the in-plane
inversion symmetry $\vare(x,y,z) = \vare(-x,-y,z)$ or $C_2$ 
symmetry (rotation by $180^\circ$ about the $z$ axis). As we have
mentioned in Sec.~\ref{S2}, this symmetry allows us to scale the BIC
and the scattering solutions, so that their $x$ and $y$ components are
${\cal PT}$-symmetric and their $z$ components are anti-${\cal
  PT}$-symmetric, as in Eq.~(\ref{gPT})~\cite{para21}. For $x$-invariant
structures, ${\bm \Phi}_*$ and ${\bm \Psi}_k$ are functions of $y$ and
$z$ only, then the ${\cal PT}$ and anti-${\cal PT}$ symmetry
conditions are
\begin{equation}
  \label{2DPT}
  \begin{bmatrix}
    \overline{P}_{x}(-y,z) \cr 
    \overline{P}_{y}(-y,z) \cr 
    -\overline{P}_{z}(-y,z) 
  \end{bmatrix}
  =
    \begin{bmatrix}
      P_{x}(y,z) \cr 
      P_{y}(y,z) \cr 
      P_{z}(y,z) 
  \end{bmatrix}
\end{equation}
where $P_x$, $P_y$ and $P_z$ are the components of ${\bm \Phi}_*$ or
${\bm \Psi}_k$ and $\overline{P}_x$ is the complex conjugate of
$P_x$.

For a generic BIC with $\alpha_* \ne 0$ and $\beta_* \ne 0$, we have 
$N_{\rm ss}=4$ and $N_{\rm wv}=2$ as before. The ${\cal PT}$ and 
anti-${\cal PT}$ symmetry  reduces the index $I_k$ from 2 to
$1$. Therefore, the codimension of the BIC is $n = 4 \times 1 -2 =
2$. 

For a BIC with $\alpha_* \ne 0$ and $\beta_* = 0$, we have $N_{\rm
  ss}=4$ and $N_{\rm wv}=1$.
Since the BIC and the scattering
solutions can be scaled such that their $x$ components are pure
imaginary and their $y$ and $z$ components are real, we have
\begin{equation}
  \label{scaleRI}
    \begin{bmatrix}
      P_{x}(y,z) \cr 
      P_{y}(y,z) \cr 
      P_{z}(y,z) 
    \end{bmatrix}
    =
    \begin{bmatrix}
      i Q_{x}(y,z) \cr 
      P_{y}(y,z) \cr 
      P_{z}(y,z) 
  \end{bmatrix},      
\end{equation}
where $Q_x$, $P_y$ and $P_z$ are real. The ${\cal PT}$ and 
anti-${\cal PT}$ symmetry, i.e., Eq.~(\ref{2DPT}), is obtained under a
different scaling which can be compensated by multiplying a constant
$C$ of unit magnitude to the left hand side of  Eq.~(\ref{2DPT}).
Therefore,
\begin{equation}
  \label{Cp1m1}
    \begin{bmatrix}
      i Q_{x}(y,z) \cr 
      P_{y}(y,z) \cr 
      P_{z}(y,z) 
    \end{bmatrix}
    = C
    \begin{bmatrix}
      -i Q_{x}(-y,z) \cr 
      P_{y}(-y,z) \cr 
      -P_{z}(-y,z) 
    \end{bmatrix}. 
  \end{equation}
The above implies that $C$ is real and can only be $1$ or $-1$. Among
the four scattering solutions, two have $C=1$ and the other two have
$C=-1$. If the scattering solution has a different value of $C$ with
the BIC, then the index is $I_k = 0$. Therefore, $n = 1+1+0 + 0 -1 = 1$. 

The BIC with $\alpha_* = 0$ and $\beta_* \ne 0$ is a scalar mode, thus
$N_{\rm ss} = 2$. Using the scaling for ${\cal PT}$ and anti-${\cal
  PT}$ symmetry, we have $I_k = 1$. Therefore $n = 2\times 1 -1 = 1$.

The BIC with $\alpha_* = \beta_* = 0$ is a scalar standing wave. If it
is generic, it has the opposite even/odd parity in $y$ with the
scattering solutions of the same polarization. Therefore, the BIC 
is symmetry-protected and $n = 0$.

In Refs.~\cite{para20,para21}, detailed derivations are given for
the results listed in the last two rows of Table~\ref{tab1}.
For the two cases considered in this section (no symmetry and
reflection symmetry in $y$), similar derivations can be worked
out following the brief discussions above. In Table~\ref{tab1}, the
results are shown only for BICs satisfying condition
(\ref{1rad}). Using the general formula (\ref{formula}), it is not
difficult to calculate the codimension $n$ for BICs that fail to satisfy
(\ref{1rad}).

\section{Numerical results}
\label{S4}

\begin{figure}[htbp!]
	\centering 
	\subfloat{	\includegraphics[width=0.5\linewidth]{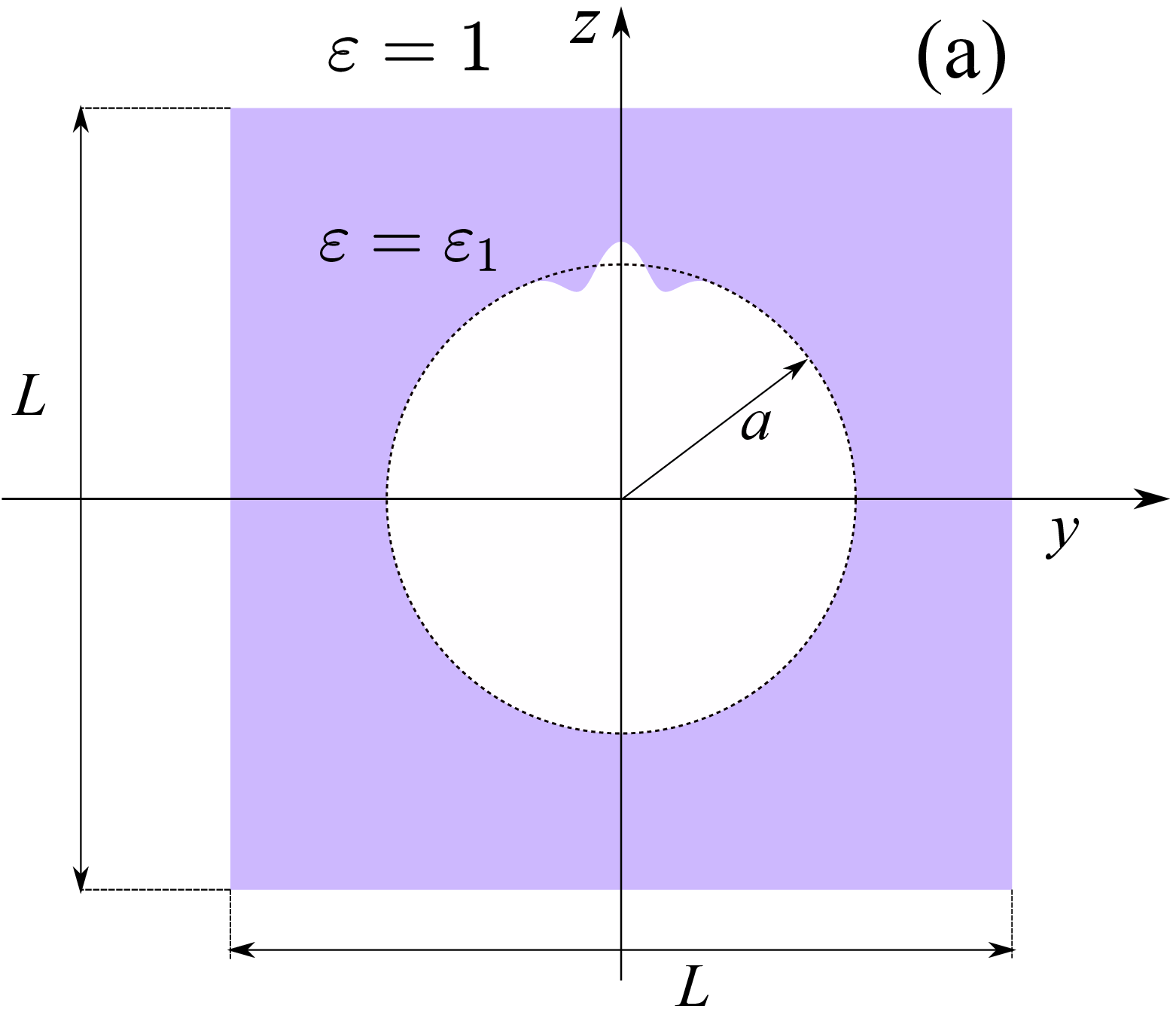}}~\subfloat{	\includegraphics[width=0.5\linewidth]{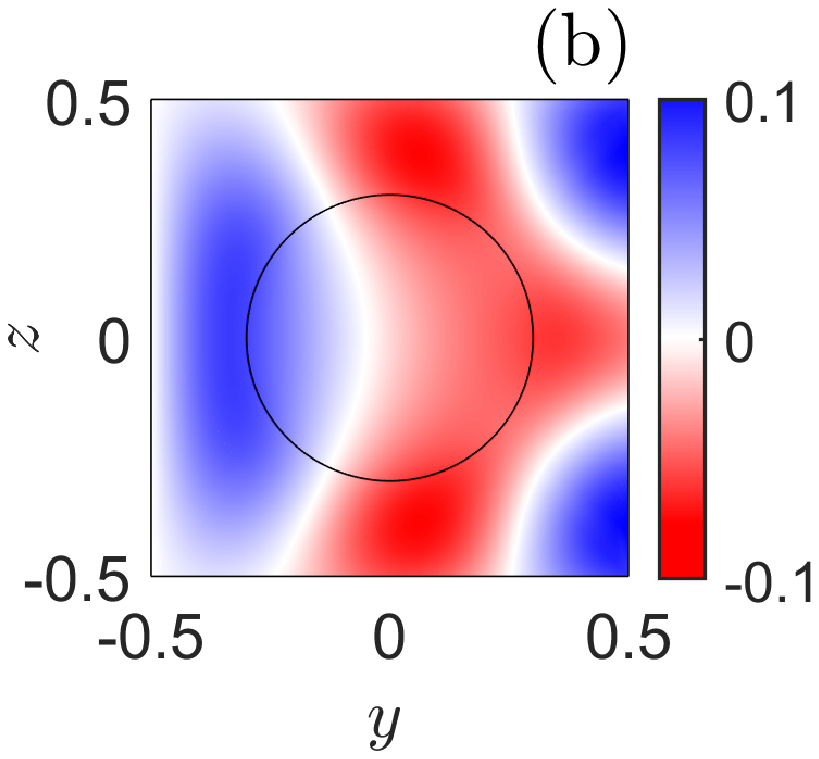}}\\
	\subfloat{	\includegraphics[width=0.5\linewidth]{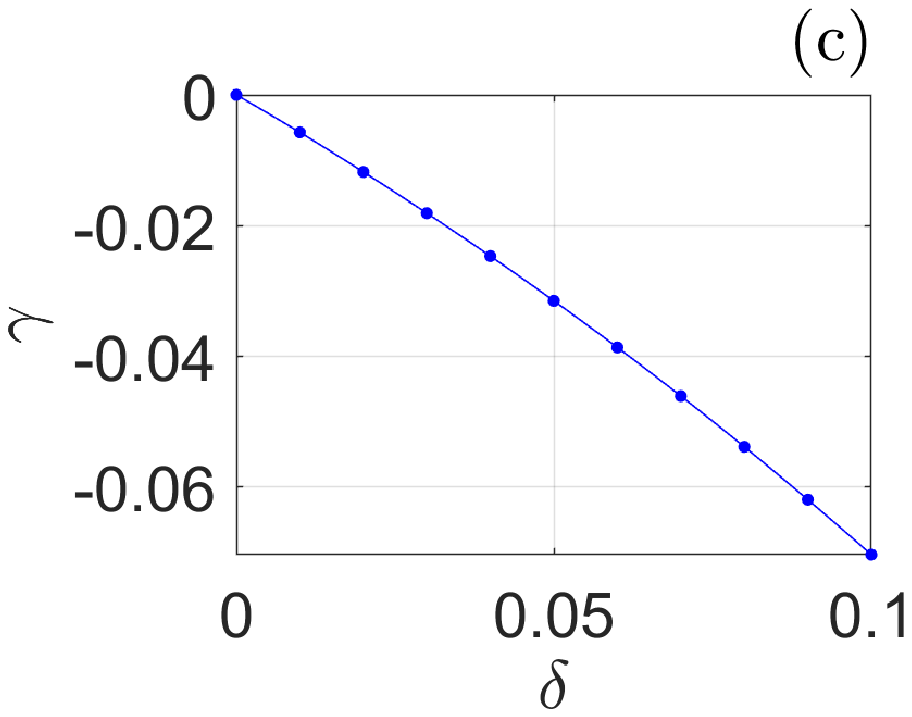}}~\subfloat{	\includegraphics[width=0.5\linewidth]{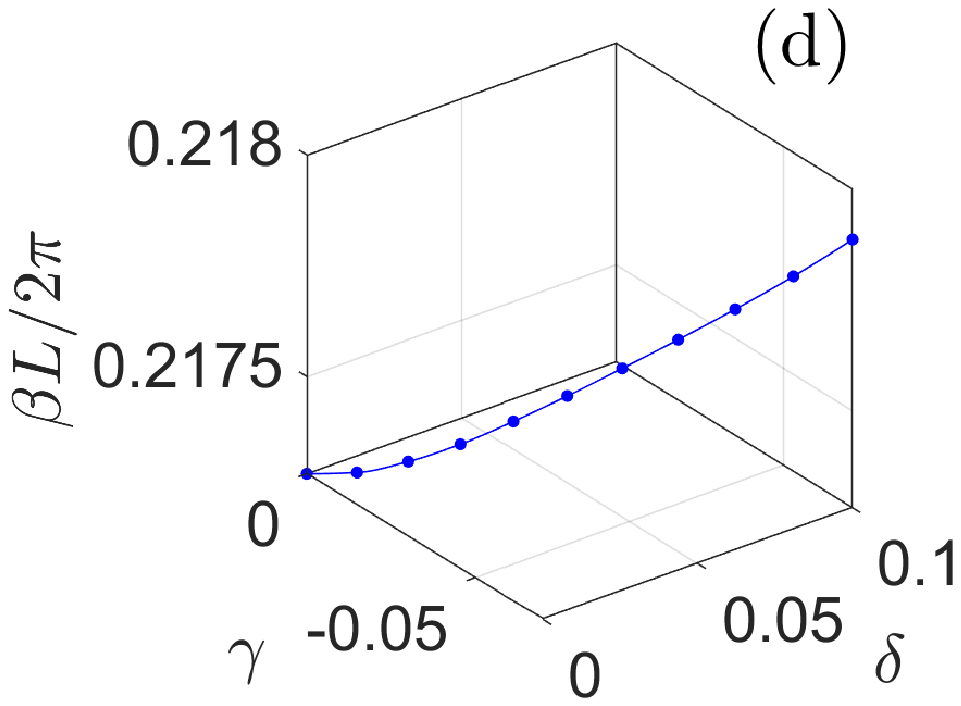}}\\
	\subfloat{	\includegraphics[width=0.5\linewidth]{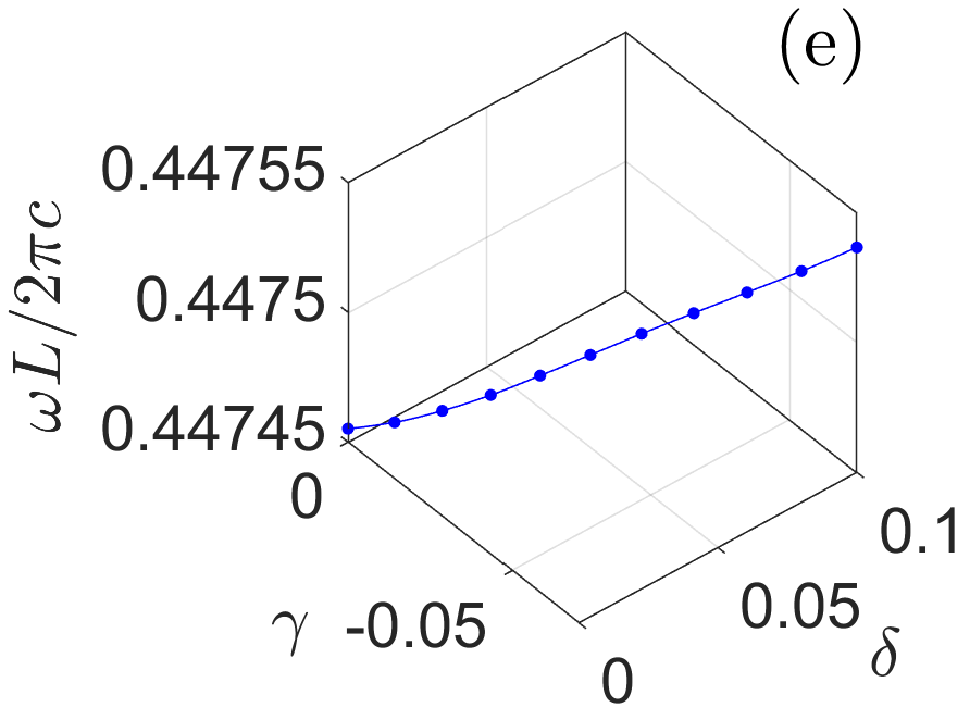}}~\subfloat{	\includegraphics[width=0.5\linewidth]{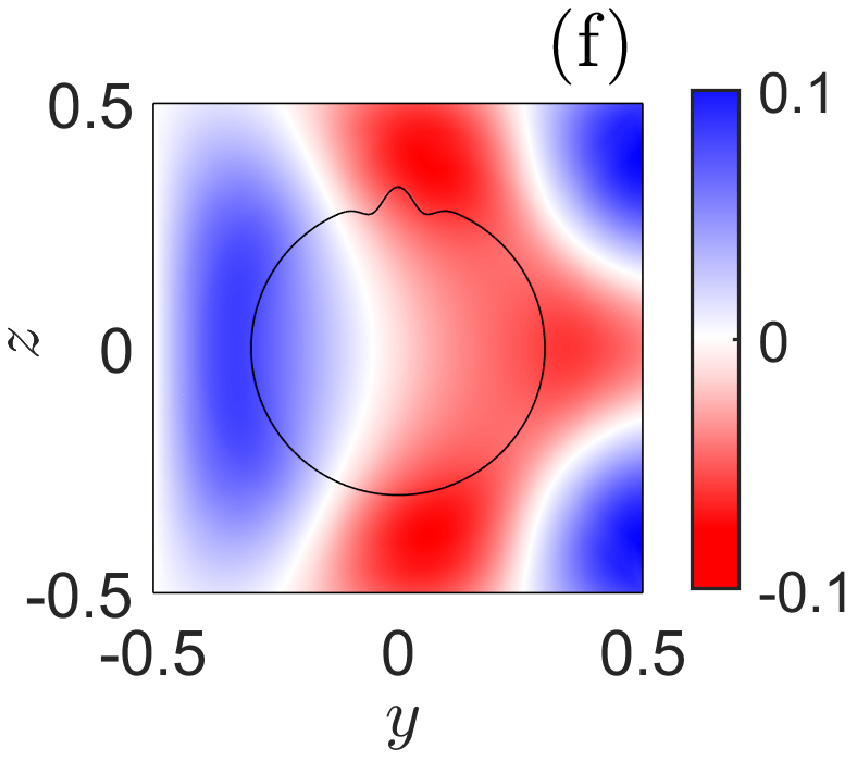}}
	\caption{(a) One period of the slab with an array of distorted
          air holes (for $\delta = 0.1$ and $\gamma = -0.1$) breaking
          the reflection symmetry in $y$. (b) Electric field
          pattern of a propagating BIC in a slab with circular air
          holes of radius $a = 0.3L$. (c) Parameter curve of the
          BIC. (d) and (e) Wavenumber $\beta$ and frequency $\omega$
          of the BIC depending on parameters $\delta$ and $\gamma$. (f) Electric field
          pattern of the BIC at $\delta = 0.1$ and $\gamma \approx -0.07$.}
	\label{Fig1}
      \end{figure}

To validate the theory developed in the previous sections, we use a
highly accurate boundary integral equation method~\cite{jcp12} to
analyze two periodic structures with broken up-down mirror
symmetry. The first example is a slab with a periodic array of air
holes shown in Fig.~\ref{Fig1}(a). 
The thickness and dielectric constant of the slab are $2d=L$ and
$\vare_1 = 11.56$, respectively, where $L$ is period in the $y$
direction. The medium above and below the slab (i.e. for $|z| >
d$) is air. The boundary of the air hole is perturbed from a
circle of radius $a$, such that
the reflection symmetry in $y$ is preserved and the reflection
symmetry in $z$ is broken. The perturbed boundary consists of two
independent real parameters $\delta$ and $\gamma$ which 
correspond to the height (relative to $a$) of the bump and the dents
shown in Fig.~\ref{Fig1}(a). The precise formula of the boundary is
\begin{equation}
  \label{pert2}
	y = a \rho(s) \cos (s),  \ z = a\rho(s) \sin(s),\quad 0\leq s < 2\pi,
\end{equation}
where $\rho(s) =   1+ \delta g(s)
     + \gamma [ g(s-0.2)
     + g(s+0.2) ]$, and 
     $g(s) = \exp (-100 |s -\pi/2|^2)$. 
For $a=0.3L$ and $\delta = \gamma = 0$, the air holes are circular and the
structure has an $E$-polarized propagating BIC with $\alpha_*=0$,  $\beta_* =
0.21728(2\pi/L)$ and $\omega_* = 0.44746(2\pi c/L)$. The electric
field distribution (real part of the $x$ component) of the BIC is
shown in  Fig.~\ref{Fig1}(b). 

\begin{figure}[ht!]
	\centering 
	\includegraphics[width=0.9\linewidth]{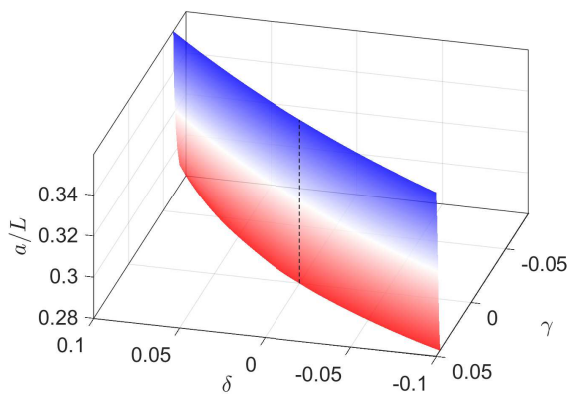}
	\caption{A  surface in the 3D space of $\delta$,
          $\gamma$ and $a$, for a BIC in a slab with an array of distorted
          air holes breaking the reflection symmetry in $y$.}
	\label{Fig2}
\end{figure}

According to the theory of Sec.~\ref{S3}, for perturbations that
preserve the reflection symmetry in $y$, the scalar propagating BIC
should have codimension $n=1$. Therefore, for the fixed $a=0.3L$,
the BIC should form a  curve in the $\delta$-$\gamma$ plane. 
This is confirmed by the numerical result shown in Fig.~\ref{Fig1}(c). The Bloch wavenumber $\beta$ and frequency $\omega$ of this BIC are
shown as functions of $\delta$ and $\gamma$ in
Figs.~\ref{Fig1}(d) and \ref{Fig1}(e), respectively. For $\delta=0.1$, the BIC is obtained 
with $\gamma \approx -0.07$ for $\beta =  0.21789(2\pi/L)$ and $\omega =
0.44754(2\pi c/L)$. Its electric field distribution (the real part of
$E_x$)  is shown in Fig.~\ref{Fig1}(f). Moreover, since its codimension is 1,
this BIC should form a surface in the space of three parameters. To
illustrate this, we simply allow $a$ to be the third parameter. In
Fig.~\ref{Fig2}, 
we show a surface of this BIC in the 
$\delta$-$\gamma$-$a$ space for  $0.28 \le a/L \le 0.36$. 
Notice that this surface includes the vertical axis at $\delta = \gamma = 0$.
It simply means that if the air holes are circular, the BIC is robust
with respect to changes in the radius.

\begin{figure}[htbp!]
	\centering 
	\subfloat{	\includegraphics[width=0.5\linewidth]{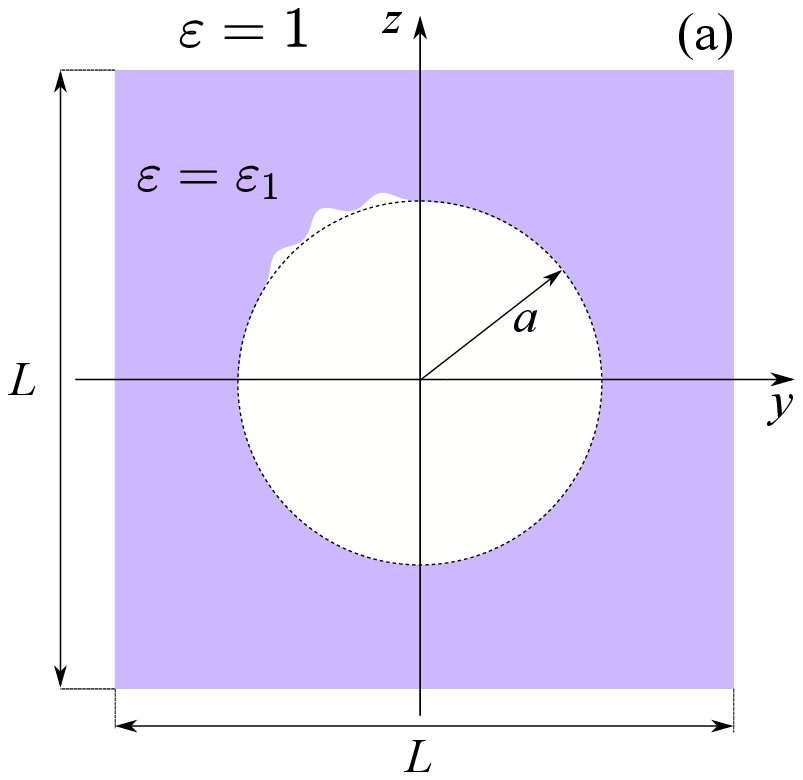}}~\subfloat{	\includegraphics[width=0.5\linewidth]{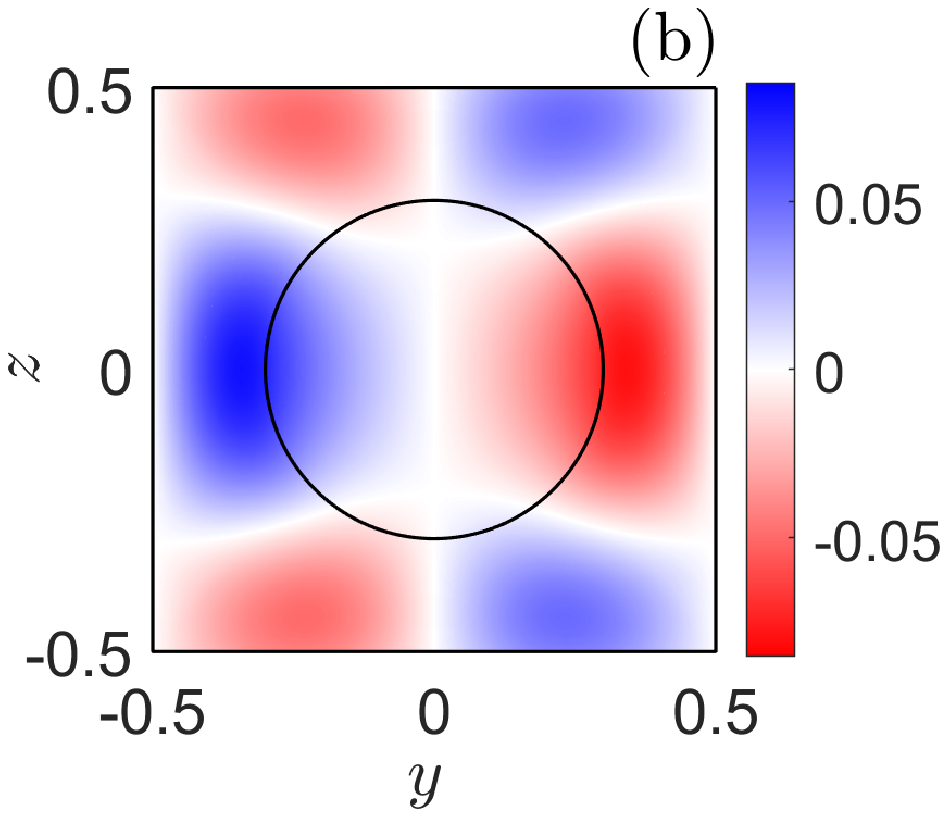}}\\
	\subfloat{	\includegraphics[width=0.5\linewidth]{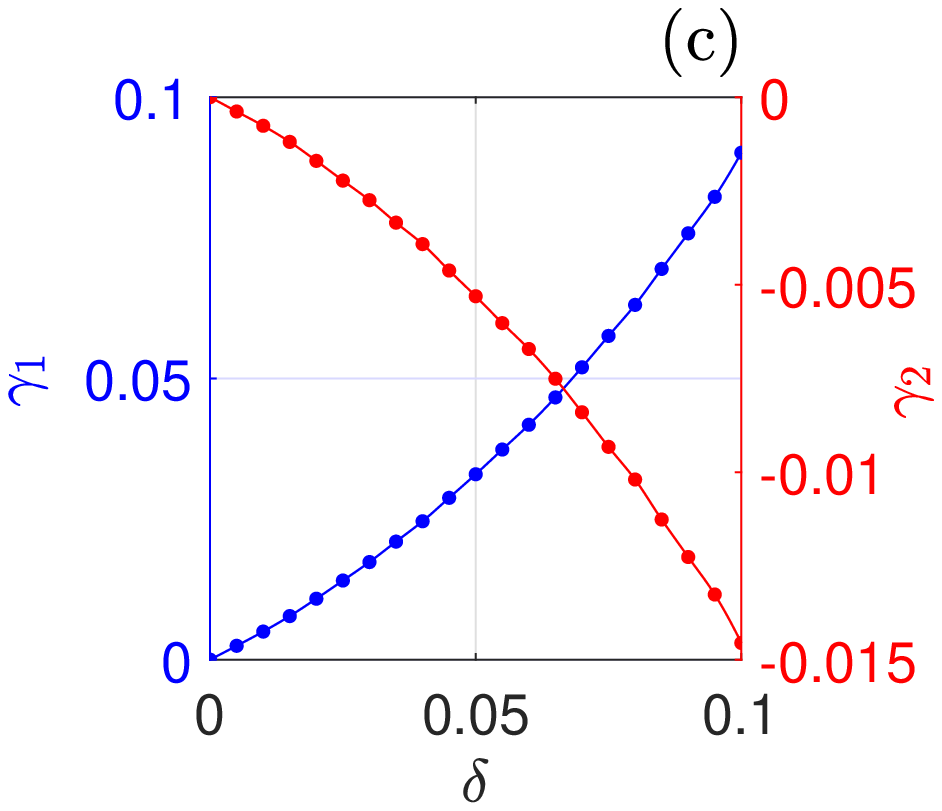}}~\subfloat{	\includegraphics[width=0.5\linewidth]{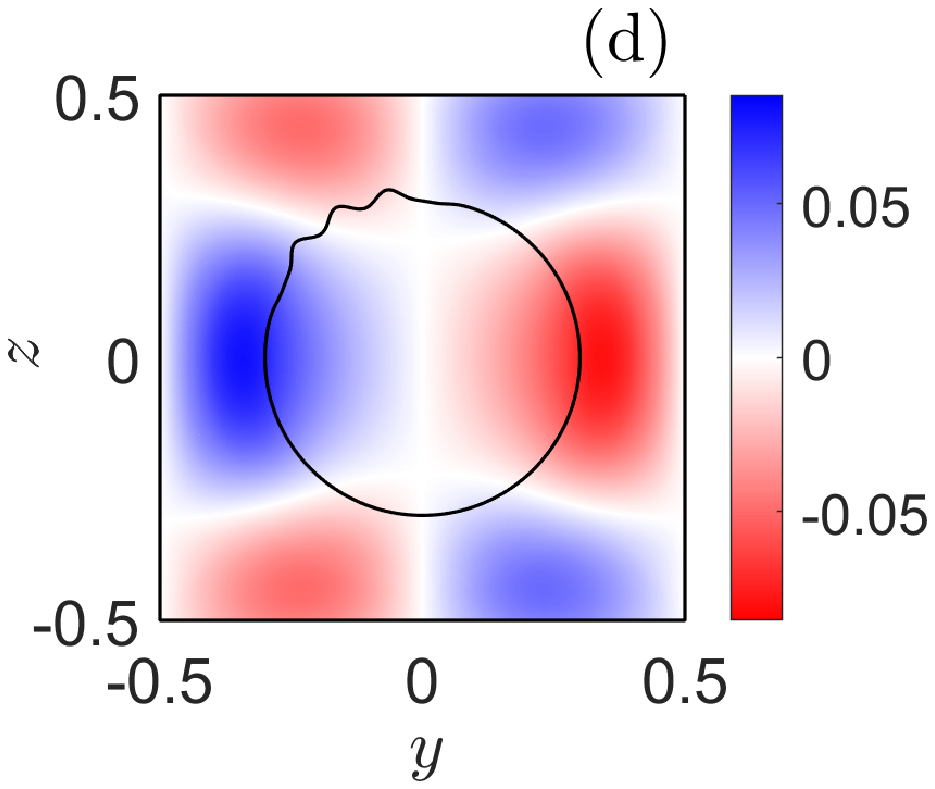}}
	\caption{(a) One period of a slab with an array of distorted
          air holes breaking both reflection symmetries in $y$ and $z$. (b)
          Electric field pattern of a standing wave in a slab with
          circular air holes of radius $a = 0.3L$. (c) Tuned
          parameters $\gamma_1$ and $\gamma_2$ as functions of
          $\delta$ for a standing wave in the slab with distorted air holes.
          (d) Electric field pattern of the standing wave for
          $\delta=0.1$ and corresponding values of $\gamma_1$ and $\gamma_2$. }
	\label{Fig3}
\end{figure}

The second example is also a slab with a periodic array of air holes,
but the boundaries of the air holes are perturbed to break both
reflection symmetries in  $y$ and in $z$.  The slab has the same
thickness $2d=L$  and the same dielectric constant $\vare_1 = 11.56$. The
boundary of the air hole is also given by Eq.~(\ref{pert2}), but
$\rho(s)$ is now given by 
\begin{eqnarray*}
  \rho(s)
  = 1 + \delta g(s)  + \sum_{j=1}^3
  \gamma_j \left[ g \left(s- \frac{j\pi}{10} \right) +
  g \left(s + \frac{j\pi}{10} \right)   \right],
\end{eqnarray*}
and $g(s)$ is given by $g(s)  = \exp(-100 |s-2\pi/3|^2)$.
Since the codimension of a scalar propagating BIC in a periodic 
structure without symmetry is $n=3$, we have introduced three 
tunable parameters $\gamma_j$ for $1 \le j \le 3$. However, if the 
BIC is a standing wave (in such a structure without symmetry), the
codimension is $n =2$.
In Fig.~\ref{Fig3}(a),
we show a unit cell of the periodic 
structure for $a = 0.3L$,  $\delta = 0.1$,  $\gamma_1 =  0.06456$, 
$\gamma_2 =  -0.00524$,  and $\gamma_3 = -0.006$. 

In the structure with circular air holes of radius $a = 0.3L$, there is a BIC (a
symmetry-protected standing wave)  with frequency $\omega_*  = 0.52044(2\pi 
c/L)$. Its electric field is shown in Fig.~\ref{Fig3}(b). For
$\delta>0$, we keep $\gamma_3 = 0$ and tune $\gamma_1$ and $\gamma_2$
to preserve this standing wave. 
The resulting $\gamma_1$ and $\gamma_2$ are shown in
Fig.~\ref{Fig3}(c) as functions of 
$\delta$ for $0\leq \delta\leq 0.1$. For $\delta = 0.1$, the 
frequency of the BIC is $\omega =  0.52053(2\pi c/L)$ and its electric field 
is shown in Fig.~\ref{Fig3}(d).

\begin{figure}[htbp!]
 	\centering 
 	\includegraphics[width=\linewidth]{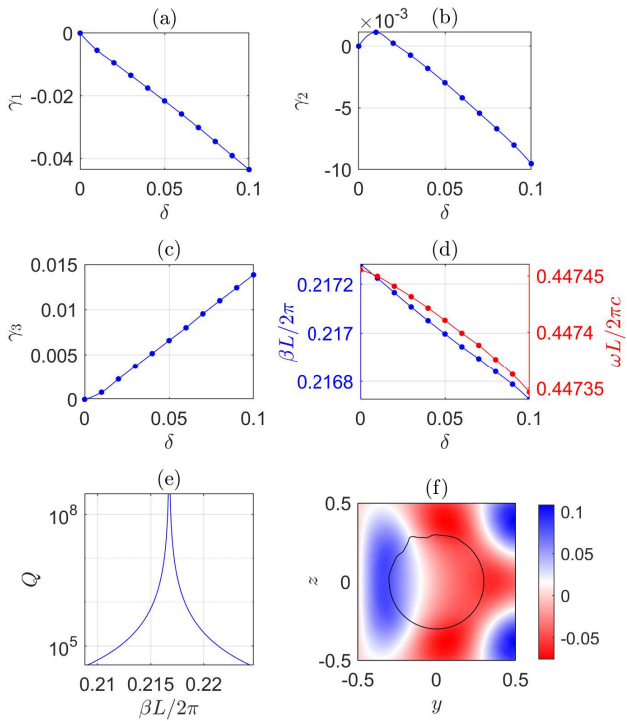}
 	\caption{ (a), (b) and (c) Tuned parameters $\gamma_1$,
          $\gamma_2$ and $\gamma_3$ as functions of
          $\delta$ for a propagating BIC in a slab with distorted air
          holes breaking both reflection symmetries in $y$ and $z$. 
          (d)  Wavenumber $\beta$ and frequency $\omega$ of the BIC as functions
          of  $\delta$.
          (e) $Q$ factor of the resonant modes near the BIC for $\delta=0.1$.
          (f) Electric field pattern of the BIC for $\delta = 0.1$. } 
 	\label{Fig4}
 \end{figure}

To preserve a propagating BIC in such a periodic structure without
symmetry, three tuning parameters are needed. Starting from the same
propagating BIC studied in the first example (for circular air hole
with radius $a=0.3L$), we calculate $\gamma_1$, $\gamma_2$ and
$\gamma_3$ to follow the BIC as $\delta$ is increased from $0$, and
show the numerical results in Fig.~\ref{Fig4}.
The three parameters as shown as functions of $\delta$ in
Figs.~\ref{Fig4}(a), \ref{Fig4}(b) and \ref{Fig4}(c), respectively.
The frequency and Bloch wavenumber of
 the BIC are shown in Fig.~\ref{Fig4}(d). For $\delta
 = 0.1$, the BIC   has $\beta= 0.21673 (2\pi/L)$ and 
 $\omega  = 0.44735 (2\pi c/L)$, and is obtained at $\gamma_1 =
 -0.04357$, $\gamma_2 =  -0.00952$,  and $ \gamma_3 =  0.01385$.
 In Fig.~\ref{Fig4}(e), we show the quality factor, $Q =
 -0.5 \RE(\omega)/\IM(\omega)$, of nearby resonant modes.
 The electric field of the BIC (the real part of $E_x$) is shown in Fig.~\ref{Fig4}(f).

\section{Conclusion}
\label{S5}

For theoretical interest and practical applications, it is important
to find out what will happen to a BIC when the structure is
perturbed. Although it typically becomes a resonant state of finite
$Q$-factor, the BIC may persist either because it is robust against a
class of perturbations, or the perturbation contains a sufficient
number of tunable parameters. We have proposed a general formula for the
minimum number $n$ of tunable parameters needed to preserve a
generic nondegenerate BIC, and calculated $n$ for BICs in 2D structures
that are invariant in one spatial direction and periodic in
another. The integer $n$ is only defined when conditions 
(typically some symmetry) on structural perturbations are properly
specified. For $n=0$ and $n\ge 1$, the BIC is robust and nonrobust,
respectively. A larger $n$ means that the BIC is difficult to find. A
different point of view is to consider structures depending on a
number of parameters. The set
of parameter values at which a BIC exists form a geometric object in the
parameter space, and the codimension of that geometric object is exactly $n$.

Although we have only verified our results for BICs in 2D 
structures with a single periodic direction, the general formula
(\ref{formula}) is proposed for all generic nondegenerate electromagnetic BICs. 
In fact, we believe the general formula for $n$ is valid for any
classical or quantum wave system. However, the current theory is only
applicable to generic BICs that guarantee $n$ can be defined. Under a
structural perturbation, a generic BIC either becomes a resonant state
or continues its existence with slightly different frequency and
wavevector.  In contrast, a non-generic BIC has the additional
possibility of splitting into two or more generic BICs. 
It is worthwhile to extend our theory to non-generic BICs, since they
have additional interesting properties and  valuable applications. 
% \vspace{0.3cm}

\section*{Acknowledgments}
L.Y. acknowledges support from Natural Science
Foundation of Chongqing, China (Grant No. cstc2019jcyj-msxmX0717).
W.L.  is partially supported by NSFC Grant 12174310 and by the NSF of Zhejiang
Province for Distinguished young scholars
LR21A010001.   Y.Y.L. acknowledges support from the Research Grants
Council of Hong Kong Special Administrative Region, China (Grant No. CityU
11307720).

% \bibliography{MyBib}%

\end{document}